\documentclass[prl,twocolumn,showpacs]{revtex4}

\usepackage{graphicx}
\usepackage{amsmath}

\newcommand{\cH}{{\cal H}}
\newcommand{\cL}{{\cal L}}
\newcommand{\cS}{{\cal S}}
\newcommand{\Eb}{E_{\rm b}}
\newcommand{\eg} {{e.g., }}
\newcommand{\Es}{E_{\rm s}}
\newcommand{\half} {\frac{1}{2}}
\newcommand{\ie} {{i.e., }}
\newcommand{\pd} {\partial}

\begin{document}

%===================================================================
\title{Compression Induced Folding of a Sheet: An Integrable System}
%===================================================================

\author{Haim Diamant} 
\email{hdiamant@tau.ac.il} 
\affiliation{Raymond \& Beverly Sackler School of Chemistry, Tel Aviv
University, Tel Aviv 69978, Israel}

\author{Thomas A.\ Witten} 
\email{t-witten@uchicago.edu}
\affiliation{Department of Physics and James Franck Institute,
University of Chicago, Chicago, Illinois 60637, USA}

\date{\today}

\begin{abstract}
  The apparently intractable shape of a fold in a compressed elastic
  film lying on a fluid substrate is found to have an exact solution.
  Such systems buckle at a nonzero wavevector set by the bending
  stiffness of the film and the weight of the substrate fluid.  Our
  solution describes the entire progression from a weakly displaced
  sinusoidal buckling to a single large fold that contacts itself.
  The pressure decrease is exactly quadratic in the lateral
  displacement.  We identify a complex wavevector whose magnitude
  remains invariant with compression.
\end{abstract}


\maketitle
%------------------------------------------------

Composite structures, containing a fluid substrate covered by a thin
rigid layer, are commonly found in biological tissues and synthetic
coatings. Unlike a freely suspended sheet, a supported layer has an
intrinsic length scale arising from the competition of bending and
substrate energy. Thus, \eg a compressed sheet floating on a fluid
buckles at a wavelength $\lambda=2\pi[B/(\rho g)]^{1/4}$, $B$ being
the bending stiffness, $\rho$ the fluid mass density and $g$ the
gravitational acceleration
\cite{Milner,CerdaMaha,VellaMaha,Huang2007,ZhangWitten,BennyPRE,Huang2010,Vella2010}.
An analogous argument holds for an elastic foundation \cite{Thompson}.

In the elastic case, it has long been recognized that this extended
periodic wrinkling is always unstable against localized folding for a
sufficiently large system
\cite{Hunt1993,LeeWaas1996,Audoly2008,Reis2009,Brau2011}.  With a
fluid substrate the same instability obtains
\cite{LukaCerda,Leahy2010,Holmes2010,DWpreprint,Audoly2011}.  
%Thus, even infinitesimal
%wrinkles are unstable against the accumulation of deformation in a
%finite domain, and large-amplitude nonlinear effects are inevitably
%relevant. 
Such fold localization has been observed in diverse fluid-supported
films\,---\,from monolayers and trilayers of nanometer-sized gold
particles \cite{LukaCerda,Leahy2010}, through submicron-thick
polymer films \cite{Holmes2010}, to $10$-$\mu$m-thick plastic sheets
\cite{LukaCerda}. It has been suggested that the localized folds,
observed in certain surfactant monolayers at the water--air interface
upon sufficiently fast compression (albeit apparently without prior
wrinkling)
\cite{Lee1998,Gopal2001,Ybert2002,Fischer2005,jpc06,Longo2009}, and
believed to be important for the function of lungs \cite{Lee2008}, may
be a manifestation of the same phenomenon \cite{LukaCerda,DWpreprint}.

The shape of the fold beyond infinitesimal amplitude has only been
known numerically \cite{LukaCerda,DWpreprint}.  The numerical studies
showed puzzling regularities.  For example, the surface pressure
appeared to vary exactly quadratically with the displacement.  Here we
account for these regularities by solving the nonlinear equation for
the fold shape exactly. This allows a much deeper analysis of the
phenomenon, including its large-deformation limit and the point of
self-contact, which are central to the folding observed in
experiments. In a broader context, the current work adds an item to
the precious collection of exactly solvable nonlinear physical
problems.

\begin{figure}[tbh]
%\vspace{0.3cm}
\centerline{\resizebox{0.4\textwidth}{!}
{\includegraphics{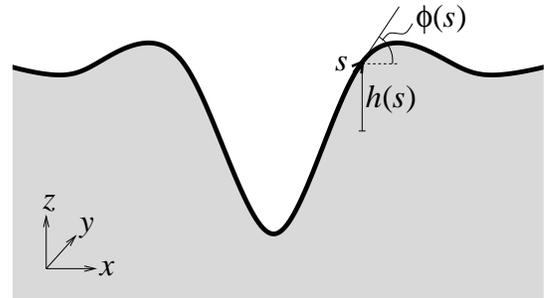}}}
\caption[]{Schematic view of the system and its parametrization.}
\label{fig_scheme}
\end{figure}

Consider a thin incompressible elastic sheet of length $L$, width $W$,
and bending modulus $B$. The sheet is uniaxially compressed along the
$x$ direction and assumed to deform in the $xz$ plane while remaining
uniform along the $y$ direction; see Fig.\ \ref{fig_scheme}. Due to
incompressibility, the configuration of the sheet is completely
defined by the profile of the angle, $\phi(s)$, that the local tangent
to the sheet makes with the $x$ axis at arclength $s$.  Alternatively,
we can define a height profile, $h(s)$, where $\dot{h}=\sin\phi$ (the
dot denoting an $s$-derivative).  The region $z<h$ is occupied by a
fluid of mass density $\rho=K/g$. We focus here on localized
deformations, and therefore let $L\rightarrow\infty$ and set
$\phi=\dot\phi=h=0$ at $s\rightarrow\pm\infty$.

The energy $E=\Eb+\Es$ contains contributions from bending,
$\Eb=(WB/2)\int_{-\infty}^\infty ds \dot{\phi}^2$, and from the
substrate energy, $\Es=(WK/2)\int_{-\infty}^\infty ds h^2\cos\phi$. The
displacement along the direction of compression is
\begin{equation}
  \Delta = \int_{-\infty}^\infty ds (1-\cos\phi), 
\label{Delta}
\end{equation}
and is related to the pressure by $P=dE/d\Delta$. For brevity we use
hereafter units where $B=K=1$, \ie we rescale energy by $B$ and length
by $(B/K)^{1/4}$, and also let $W=1$. The pressure $P$ is scaled by
$(BK)^{1/2}$.

Invoking a dynamical analogy where $s$ stands for time, we look for
the stable configuration by minimizing the action $\cS =
\int_{-\infty}^{\infty} ds \cL(\phi,h,\dot\phi,\dot{h})$,
\begin{equation}
  \cL = \half\dot\phi^2 + \half h^2\cos\phi - P(1-\cos\phi)
  - Q(s)(\sin\phi-\dot{h}),
\label{S}
\end{equation}
where $P$ and $Q(s)$ are Lagrange multipliers replacing, respectively,
the global constraint on $\Delta$ [Eq.\ (\ref{Delta})] and the local
one on the relation between $h$ and $\phi$. (In the case of an elastic
foundation, the hydrostatic $(h^2/2)\cos\phi$ term is replaced by
$h^2/2$ \cite{Thompson}.) We identify the conjugate momenta as
$p_\phi=\pd\cL/\pd\dot\phi=\dot\phi$ and $p_h=\pd\cL/\pd\dot{h}=Q$,
and use them to obtain the Hamiltonian,
$\cH=p_\phi\dot\phi+p_h\dot{h}-\cL$. Since $\cL$ has no explicit
dependence on $s$ (the sheet is translation-invariant), $\cH$ is a
constant of motion,
\begin{equation}
  \cH = \half p_\phi^2 + p_h\sin\phi - \half h^2\cos\phi + P(1-\cos\phi) = 0,
\label{H}
\end{equation}
where the last equality follows from the boundary conditions at
$s\rightarrow\pm\infty$.  Equation (\ref{H}) has the consequence that,
wherever the sheet is horizontal ($\phi=0$), we have $|p_\phi|=|h|$,
which leads to the geometrical constraint,
\begin{equation}
  \phi=0:\ \ \ |\dot\phi| = |h| = |\ddot{h}|.
\label{horizontal}
\end{equation}

Hamilton's equation, $\dot{p}_\phi=-\pd\cH/\pd\phi$, yields the following
equation of motion:
\begin{equation}
  \ddot\phi + (h^2/2 + P)\sin\phi + p_h\cos\phi = 0.
\label{motion1}
\end{equation}
%
%We note that without the hydrostatic terms ($h=p_h=0$) this is the
%equation of motion of a physical pendulum. 
Eliminating $p_h$ from Eqs.\ (\ref{H}) and (\ref{motion1}) and
differentiating the resulting equation with respect to $s$, we get
\begin{equation}
  \dddot\phi + (\dot\phi^2/2 + P)\dot\phi + h = 0.
\label{motion2}
\end{equation}
Equation (\ref{motion2}) coincides with Euler's {\it elastica} problem
\cite{Thompson,CerdaMaha2005,WittenRMP}. It expresses the balance of
normal forces on an infinitesimal section of the sheet.  The last
term, which usually corresponds to an external normal force
\cite{CerdaMaha2005}, arises here from hydrostatic pressure.  Another
differentiation yields the equation in terms of $\phi$ alone,
\begin{equation}
  \ddddot\phi + [(3/2)\dot\phi^2 + P]\ddot\phi + \sin\phi = 0.
\label{motion3}
\end{equation}
Continuing to assume that $\phi$ and its derivatives vanish at
infinity, we integrate Eq.\ (\ref{motion3}) once to get
\begin{equation}
  \dddot\phi\dot\phi - \half\ddot\phi^2 + \frac{3}{8}\dot\phi^4
  + \half P\dot\phi^2 + 1-\cos\phi = 0.
\label{motion4}
\end{equation}

\begin{figure*}[tbh]
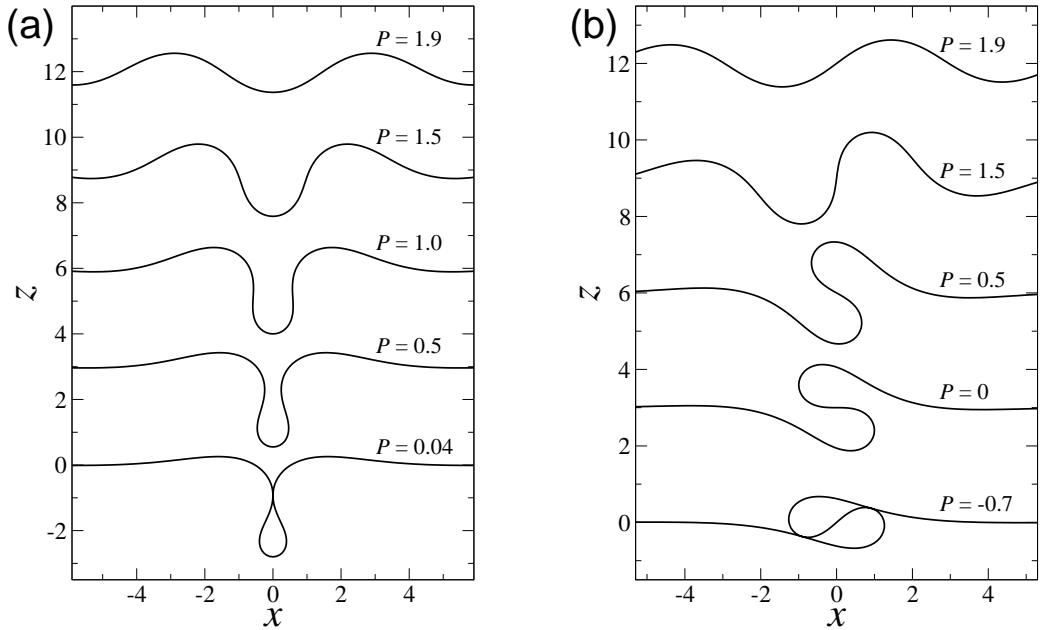

%  \vspace{0.6cm} 
  \centerline{\resizebox{0.35\textwidth}{!}
    {\includegraphics{fig2a.eps}} %}\vspace{1.1cm}
  %\centerline{
  \hspace{1cm}
  \resizebox{0.35\textwidth}{!}
    {\includegraphics{fig2b.eps}}}
\caption{Symmetric (a) and antisymmetric (b) configurations of the sheet in the 
  $xz$ plane as a function of decreasing pressure (increasing
  displacement) from a point close to the instability threshold
  ($P_{\rm c}=2$) down to self-contact. The curves are vertically shifted by
  3 from one another for clarity.}
\label{fig_folds}
\end{figure*}

At first glance, the nonlinear Eq.\ (\ref{motion4}) does not seem
likely to lend itself to a closed-form solution. Inspection of the
equations above reveals, on the other hand, that they readily yield
the profile and all of its derivatives at $s=0$. Let us specialize,
for instance, to a symmetric deformation about a downward-pointing
fold at the origin [as in Fig.\ \ref{fig_folds}(a)], where
$\phi(0)=\ddot\phi(0)=0$. We can then apply Eqs.\ (\ref{motion2}) and
(\ref{motion4}) at $s=0$ and, thanks to the geometrical condition of
Eq.\ (\ref{horizontal}), solve for $\dot\phi(0)$ and $\dddot\phi(0)$,
\begin{equation}
  \dot\phi(0) = -h(0) = 2(2-P)^{1/2},
\label{h0}
\end{equation}
and $\dddot\phi(0)=-2(3-P)(2-P)^{1/2}$. Higher derivatives are
obtained from Eq.\ (\ref{motion3}) and its successive differentiation.
The complete knowledge of the power-series at $s=0$ hints that the
problem may be integrable.

Another indication is suggested by the integrable physical-pendulum (PP)
equation, $\ddot\phi+q^2\sin\phi=0$, whose solutions are
\begin{equation}
  \phi(s) = \pm 4\tan^{-1}(A e^{\pm iqs}),
\label{wavelike}
\end{equation}
for any $q$ and $A$. It is straightforward to show, by integrating the
PP equation once and differentiating it twice, that any of its
solutions also solves Eq.\ (\ref{motion3}) for $P=q^2+q^{-2}+c$, where
$c$ is an integration constant (set hereafter to zero). Thus, Eq.\
(\ref{wavelike}) gives complex solutions to Eq.\ (\ref{motion4}), with
the specific complex wavevectors,
\begin{equation}
  q = \pm k \pm i\kappa,\ \ \ 
  k = \half (2+P)^{1/2},\  
  \kappa = \half (2-P)^{1/2},
\label{eqq}
\end{equation}
and an arbitrary amplitude $A$ (the latter following from translation
invariance).  We note that $k^2+\kappa^2=1$, independent of $P$, while
$k^2-\kappa^2=P/2$. These solutions resemble the `kink' solutions of
the sine-Gordon (SG) equation \cite{Infeld,Newell}, albeit in the
complex plane.  When linearized, they coincide with the `evanescent
wave' profile, which can be stabilized adjacent to a boundary for $P$
close to the critical pressure $P_{\rm c}$ \cite{DWpreprint}.

These findings indicate that Eq.\ (\ref{motion3}) might belong to a
hierarchy of integrable nonlinear equations \cite{Newell,Gesztesy}, in
which the PP equation is a lower-order member. A known hierarchy of
equations, referred to as the
stationary-sine-Gordon-modified-Korteweg-de-Vries hierarchy
\cite{Gesztesy}, indeed contains the stationary SG equation, the PP
equation, and Eq.\ (\ref{motion3}) as the first, second, and third
members, respectively. To our knowledge, equations in this hierarchy
beyond the PP equation have not been linked before to physical
phenomena.

To construct localized real solutions out of the complex ones given in
Eq.\ (\ref{wavelike}), we borrow a scheme from the SG problem. In
`light-cone' coordinates $[u=(x+t)/2,v=(x-t)/2]$, the SG equation,
$\pd_{xx}\phi-\pd_{tt}\phi=\pd_{uv}\phi=\sin\phi$, is invariant to the
scaling $u\rightarrow qu, v\rightarrow v/q$ by an arbitrary scale
factor $q$.  Given three known solutions of this equation, $\phi_j$
($j=0,1,2$), one can construct another solution, $\phi_3$, using the
implicit `ladder' rule \cite{Newell},
$\tan[(\phi_3-\phi_0)/4]=[(q_1+q_2)/(q_1-q_2)]\tan[(\phi_1-\phi_2)/4]$,
where $q_1$ and $q_2$ are arbitrary scale factors for $\phi_1$ and
$\phi_2$. We attempt the same procedure while substituting for $q_1$
and $q_2$ two of the specific $P$-dependent wavevectors found in Eq.\ 
(\ref{eqq}).  Choosing $\phi_0=0$, $\phi_1=4\tan^{-1}(e^{iq_1s})$, and
$\phi_2=4\tan^{-1}(e^{iq_2s})$, with $q_1=k-i\kappa$ and
$q_2=-k-i\kappa$, we obtain the odd function
$\tan(\phi_3/4)=(\kappa/k)\sin(ks)/\cosh(\kappa s)$.  Using instead
$\tan(\phi_1/4)=ie^{iq_1s}$ and $\tan(\phi_2/4)=-ie^{iq_2s}$, we get
the even counterpart.  Substitution of these two functions in Eq.\ 
(\ref{motion4}) confirms that they indeed solve it exactly. Thus, the
following are exact localized shapes of the angular profile:
\begin{eqnarray}
  \mbox{symmetric fold:}\ \  
  \phi(s) = 4\tan^{-1} \left[ \frac{\kappa\sin(ks)}{k\cosh(\kappa s)} 
    \right]~&&~~~~ \nonumber\\
  \mbox{antisymmetric fold:}\ \  
  \phi(s) = 4\tan^{-1} \left[ \frac{\kappa\cos(ks)}{k\cosh(\kappa s)}
    \right].&&
\label{solution}
\end{eqnarray}
These functions match the `breather' solutions of the SG equation
\cite{Infeld} when those are projected onto the light cone,
$(s=x=t=u,v=0)$. Due to the symmetries under reflection about the $z$
axis, reflection about the $s$ (or $x$) axis, and translation along
$s$, the functions $\pm\phi(\pm s+s_0)$, where $\phi(s)$ is either one
of the functions in Eq.\ (\ref{solution}) and $s_0$ an arbitrary
constant, are solutions as well. The existence of odd and even
solutions then follows from the aforementioned ladder rule.

Evidently, the equations simplify when $\kappa$ becomes small, as
$P\rightarrow P_{\rm c}=2$; then, \eg the symmetric fold has
$\phi\simeq 4\kappa\sin s/\cosh(\kappa s)$, which itself is
vanishingly small.  This is the regime of incipient buckling discussed
previously
\cite{Milner,CerdaMaha,ZhangWitten,DWpreprint,Audoly2011}. The
buckling is always localized, but the localization length diverges as
the threshold is approached \cite{DWpreprint,Audoly2011}.

The solution implies very simple relations among the pressure $P$, the
displacement $\Delta$, the central height $|h(0)|$, and the energies.
The decay parameter is exactly linear in the displacement,
$\kappa=\Delta/8$. Indeed, the expressions for $k$ and $\kappa$ [Eq.\ 
(\ref{eqq})] exactly match the complex wavevector obtained from a
linear analysis of the `evanescent-wave' for $P\rightarrow P_{\rm c}$
\cite{DWpreprint}. Consequently, the pressure is exactly quadratic in
$\Delta$: $P=2-\Delta^2/16$, as previously deduced in that limit
\cite{DWpreprint}. The maximum amplitude of a symmetric deformation
is $|h(0)|=\Delta/2$. The bending and substrate contributions to the
energy are $E_{\rm b}=\Delta$ and $E_{\rm s}=\Delta(1-\Delta^2/48)$.
We note that the energies and pressures are identical for the
symmetric and antisymmetric cases.

Figure \ref{fig_folds} shows the progression of the symmetric and
antisymmetric folds as the lateral displacement increases and the
pressure decreases. The configurations have been calculated from Eq.\ 
(\ref{solution}) according to the parametrization:
$x(s)=\int_0^s ds' \cos\phi(s')$, $z(s)=h(0)+\int_0^s ds'
\sin\phi(s')$, where $h(0)$ is given for the symmetric fold by Eq.\ 
(\ref{h0}) and for the antisymmetric one by $h(0)=0$.

\begin{figure}
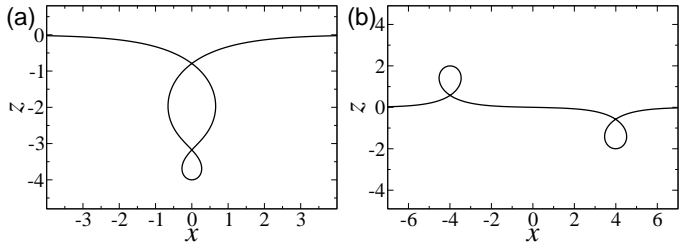

\vspace{0.6cm} 
\centerline{\resizebox{0.247\textwidth}{!}
{\includegraphics{fig3a.eps}}
\resizebox{0.247\textwidth}{!}
{\includegraphics{fig3b.eps}}}
\caption{The fully `damped' self-intersecting configurations obtained 
  for $P\rightarrow -2$. (a) Symmetric configuration for $P=-2$
  [$\phi(s)=4\tan^{-1}(s/\cosh s)$]. (b) Antisymmetric configuration
  for $P=-1.9999$; as $P\rightarrow -2$ the two loops are pushed
  toward the boundaries, leaving a flat sheet ($\phi=2\pi$) in
  between.}
\label{fig_damped}
\end{figure}

The symmetric fold is found to contact itself at a small positive
pressure, $P\simeq 0.040$ (corresponding to $\Delta\simeq 5.6\simeq
0.89\lambda$). Self-contact of the antisymmetric fold, by contrast,
requires a substantial negative pressure (\ie tension) of $P\simeq
-0.70$ ($\Delta\simeq 6.6\simeq 1.05\lambda$).  Thus, in the case of
an antisymmetric configuration the stress in the sheet vanishes prior
to self-contact. One can examine the solutions also beyond
self-contact, where they produce self-intersecting configurations
which are unphysical for a sheet. In particular, at $P=-2$ the
oscillations in $\phi$ disappear. These configurations are shown in
Fig.\ \ref{fig_damped}.

These remarkably simple yet exact results are consequences of the high
level of symmetry characteristic of integrable nonlinear problems
\cite{Newell}. The case of an elastic foundation \cite{Hunt1993} is
physically less simple, since in-plane shear forces (and not merely
normal hydrostatic ones) are exerted on the sheet. Although this
problem is found to obey the same constraint at extrema [Eq.\ 
(\ref{horizontal})], it does not exhibit the regularities described
above.

The solution presented here provides precise knowledge of shapes and
energies in a large class of wrinkling and folding systems. It enables
a new means for making precise force actuators and transducers on any
scale where uniform, thin sheets can be made, for example. The
solution improves the prospects for understanding the unstable motion
resulting from folding \cite{Fischer2005,jpc06,Longo2009} and the
observed buckling of nanoparticle monolayers into trilayers
\cite{Leahy2010}.  At the molecular scale, it provides a starting
point for quantifying the effects of compressibility and
self-attraction of surfactant monolayers, as well as the influence of
non-fluid aspects of the substrate.  More basically,
compression-induced folding appears to be a previously unrecognized
integrable solitary wave phenomenon, like the sine-Gordon chain and
the Korteweg-de Vries hydrodynamic soliton. The results above may be
used to construct more complex, multiple-fold shapes.  The fundamental
reason for the integrability of the problem remains to be
understood. It is to be hoped that this understanding will reveal a
broader class of physical systems which are integrable for the same
reason.

%------------------------------------------------
\begin{acknowledgments}
  We are indebted to Enrique Cerda for communicating early hints of
  simple features in this system, and to Ilya Gruzberg for insightful
  discussions and for recognizing our system as a member of a known
  hierarchy of integrable systems. We thank Benny Davidovitch and Leo
  Kadanoff for helpful discussions. We thank the Aspen Center for
  Physics for its hospitality during part of this work. The research
  was supported in part by the US--Israel Binational Science
  Foundation under Grant Number 2006076, and in part by the National
  Science Foundation's MRSEC Program under Award Number DMR 0820054.
\end{acknowledgments}

%------------------------------------------------
% References
%------------------------------------------------

%------------------------------------------------

\end{document}